\documentclass[a4paper]{jpsj-suppl}
\usepackage{txfonts}
\usepackage{xcolor}
\usepackage{xspace}
\usepackage{amsmath}
\usepackage{amssymb}
\usepackage{url}

\def\EeV{\ifmmode {\mathrm{Ee\kern -0.07em V}}\else
                   \textrm{Ee\kern -0.07em V}\fi}
\def\TeV{\ifmmode {\mathrm{Te\kern -0.07em V}}\else
                   \textrm{Te\kern -0.07em V}\fi}
\def\eV{\ifmmode {\mathrm{\ e\kern -0.07em V}}\else
                   \textrm{e\kern -0.07em V}\fi}
\def\gcm{\ensuremath{\mathrm{g/cm}^2}\xspace}
\def\Xmax{\ensuremath{X_\mathrm{max}}\xspace}
\def\sigmaXmax{\ensuremath{\sigma(X_\text{max})}\xspace}
\def\meanXmax{\ensuremath{\langle X_\text{max}\rangle}\xspace}
\def\Sibyll{\textsc{Sibyll2.1}\xspace}
\def\Epos{\textsc{Epos-LHC}\xspace}
\def\QgIIOld{\textsc{QGSJetII-03}\xspace}
\def\QgIINew{\textsc{QGSJetII-04}\xspace}
\def\dd{\mathrm{d}}

\newcommand{\gcmdec}{\ifmmode {\gcm/\mathrm{decade}}\else
                     {\gcm/decade}\fi\xspace}%

\title{Report of the Working Group on the Composition of
       Ultra High Energy Cosmic Rays}

\author{R.\ Abbasi$^{1}$, J.\ Bellido$^{2}$, J.\ Belz$^{1}$, V.\ de Souza$^{3}$,
W.\ Hanlon$^{1}$, D.\ Ikeda$^{4}$, J.P.\ Lundquist$^{1}$, P.\ Sokolsky$^{1}$,
T.\ Stroman$^{1}$, Y.\ Tameda$^{5}$, Y.\ Tsunesada$^{6}$, M.\ Unger$^{7,8}$,
A.\ Yushkov$^{9}$ for the Pierre Auger Collaboration$^{10}$ and the Telescope Array Collaboration$^{11}$}

\inst{
$^{1}$Department of Physics and Astronomy, University of Utah,
115 South 1400 East, Salt Lake City 84112, U.S.A.\\
$^{2}$University of Adelaide, Adelaide, S.A., Australia\\
$^{3}$Instituto de F\'isica de S\~ao Carlos, Universidade de S\~ao Paulo,
S\~ao Carlos, Brasil\\
$^{4}$ Institute for Cosmic Ray Research, University of Tokyo, Kashiwa,
Chiba, Japan\\
$^{5}$ Faculty of Engineering, Kanagawa University, Yokohama, Kanagawa, Japan\\
$^{6}$ Graduate School of Science and Engineering, Tokyo Institute of Technology,
 Meguro, Tokyo, Japan\\
$^{7}$Center for Cosmology and Particle Physics, New York University,
4 Washington Place, New York, NY 10003\\
$^{8}$Institut f\"ur Kernphysik, Karlsruher Institut f\"ur Technologie,
Postfach 3640, 76021 Karlsruhe, Germany\\
$^{9}$Institut f\"ur Physik, Universit\"at Siegen ,
Walter-Flex-Str.\ 3, 57068 Siegen, Germany\\
$^{10}$\url{http://www.auger.org/archive/authors_2014_10.html}\\
$^{11}$\url{http://www.telescopearray.org/index.php/research/collaborators}
}

\email{jose.bellidocaceres@adelaide.edu.au, belz@cosmic.utah.edu}

\abst{For the first time a proper comparison of the average depth of shower
maximum (\Xmax) published by the Pierre Auger and Telescope Array
Observatories is presented. The \Xmax distributions measured by
the Pierre Auger Observatory were fit using simulated events
initiated by four primaries (proton, helium, nitrogen and iron). The
primary abundances which best describe the Auger data were simulated
through the Telescope Array (TA) Middle Drum (MD) fluorescence and
surface detector array. The simulated events were analyzed by the TA
Collaboration using the same procedure as applied to their data. The result is a
simulated version of the Auger data as it would be observed by TA. This
analysis allows a direct comparison of the evolution of \meanXmax with
energy of both data sets. The \meanXmax measured by TA-MD is
consistent with a preliminary simulation of the  Auger data through the TA detector
and the average difference between the two data sets was found to be
$(2.9 \pm 2.7\;(\text{stat.}) \pm 18\;(\text{syst.}))$~\gcm.
}

\kword{UHECR 2014, cosmic rays, air shower, composition}

\begin{document}
\maketitle

\section{Introduction}

Composition is a central key to understand the origin of ultra-high
energy cosmic rays. Large efforts in developing new detectors and
analysis procedures have been made recently in order to improve our
knowledge about the abundance of particles arriving at Earth. At the
highest energies ($E > 10^{18}$ \eV) the depth of shower maximum
(\Xmax) is
one of the most robust observables available to infer the composition. Currently, the Pierre
Auger and the Telescope Array observatories measure \Xmax using
fluorescence detectors. Despite the use of the same detection
principle, a direct comparison of the data published by both
collaborations is not straightforward.

\begin{figure}[!t]
\centering
\includegraphics[clip, bb=0 0 144 126, width=0.48\linewidth]{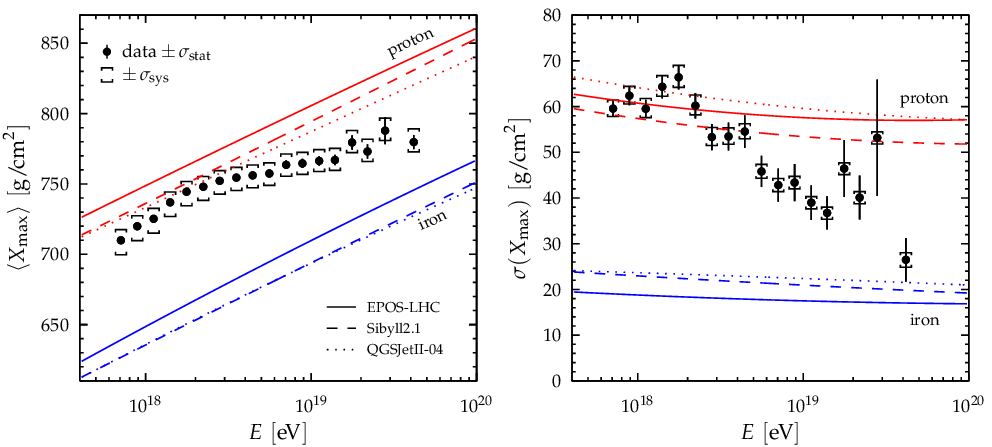}\hfill
\includegraphics[clip, bb = 6 -20 519 406, width=0.5\linewidth]{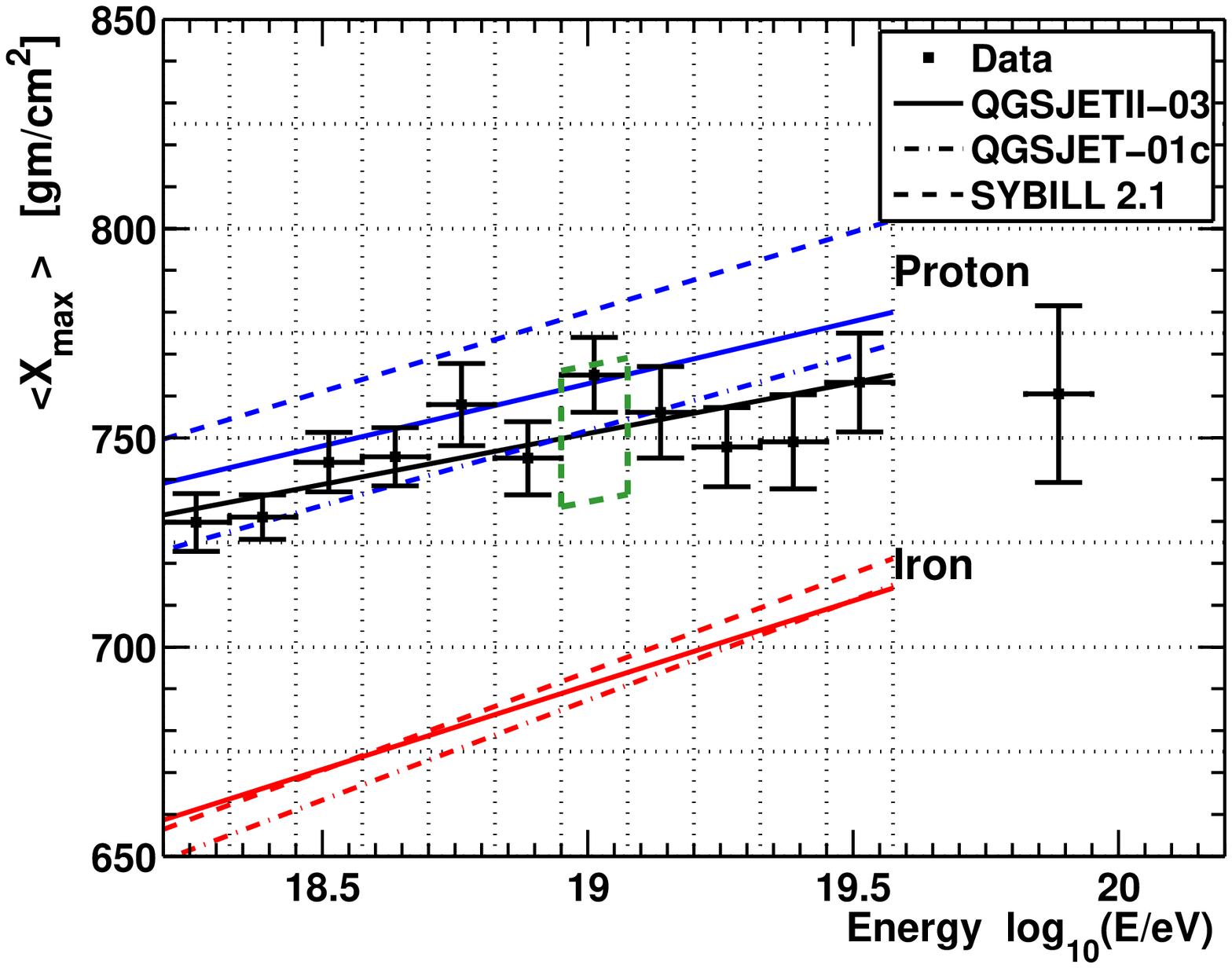}
\caption[meanXmax]{\meanXmax as measured by the Pierre Auger (left)
  and Telescope Array (right) Collaboration  \cite{bib:auger:xmax:long, Abbasi:2014sfa}. The colored lines denote predictions of air shower simulation
(note that different models are shown in the left and right panel, only \Sibyll is the same). The black line on the right panel is a straight-line fit to the TA data.}
\label{fig_meanXmax}
\end{figure}

The TA Collaboration publishes \meanXmax values obtained from
distributions folded with detector resolution and efficiency. The
interpretation of the data is made possible by the publication of the
Monte-Carlo prediction for proton and iron nuclei also folded with
detector resolution and efficiency (Fig.~\ref{fig_meanXmax}, right).
In the Auger Collaboration only certain shower geometries are selected
for sampling almost unbiased \Xmax distributions.
The corresponding \meanXmax values are presented in the left panel of Fig.~\ref{fig_meanXmax}.
 In the Auger analysis, each selected  geometry allows a wide enough
 range of \Xmax values to be observed within the fluorescence detector
 field of view boundaries. We will refer to this event selection as
 'fiducial selection'. Besides that, the Auger Collaboration published
 \sigmaXmax with detector resolution unfolded. This procedure allows
 the interpretation of the data (i.e. \meanXmax and \sigmaXmax) using Monte
 Carlo predictions without the need to fold the detector properties into the
 predictions.
 The advantage of the TA analysis is that it does not require removing as many events, since this technique does not apply a fiducial selection.\\

The work reported here is a common effort of the Auger and TA
Collaborations with the aim to provide the cosmic ray community a
 direct comparison of the \meanXmax measurements taking into account the
different approaches of each collaboration. Indirect comparisons
of TA and Auger results were published in the first report of these series~\cite{bib:uhecr:2012}. The disadvantage of indirect comparisons is that they depend on the particular hadronic interaction model that is used.
The current analysis was performed in the following way. The Auger \Xmax
distributions were fitted by a combination of four primary
nuclei (proton, helium, nitrogen, iron) using events from air shower simulations. The abundances which best fit the
Auger data were simulated through the TA-MD detector and analyzed by the TA Collaboration using
the same procedure as applied to their data. This procedure resulted in the Auger data
folded into the TA-MD detector. The Auger \meanXmax folded with TA-MD
analysis is shown in this paper in comparison to the TA-MD data as it
is usually published.

\section{Data Samples}

The analysis presented here is based on the data measured by the
Pierre Auger Observatory in the period from the 1st of December 2004 until 31st
of December 2012. All measured events were analyzed as explained in
reference~\cite{bib:auger:xmax:long}. The events were selected in
order to guarantee good measurement conditions and
a high-quality reconstruction. After that, the fiducial selection was applied. In
total 19,947 events were considered for further analysis. The \Xmax values of
these events were sampled in 18 energy bins starting at $\log{
(E/{\!\eV})} = 17.8$.

From the Telescope Array we use hybrid data collected with the MD fluorescence
telescope and surface detector array over the period from the
27th of May 2008 to the 2nd of May 2013. The reconstruction and analysis applied to
the data is described in~\cite{Abbasi:2014sfa}. The number of events which passed
all cuts is 438, for which the mean \Xmax is shown in 12 energy bins above
$\log{(E/{\!\eV})} = 18.2$.

\begin{figure}[!t]
\centering
\includegraphics[width=0.7\linewidth]{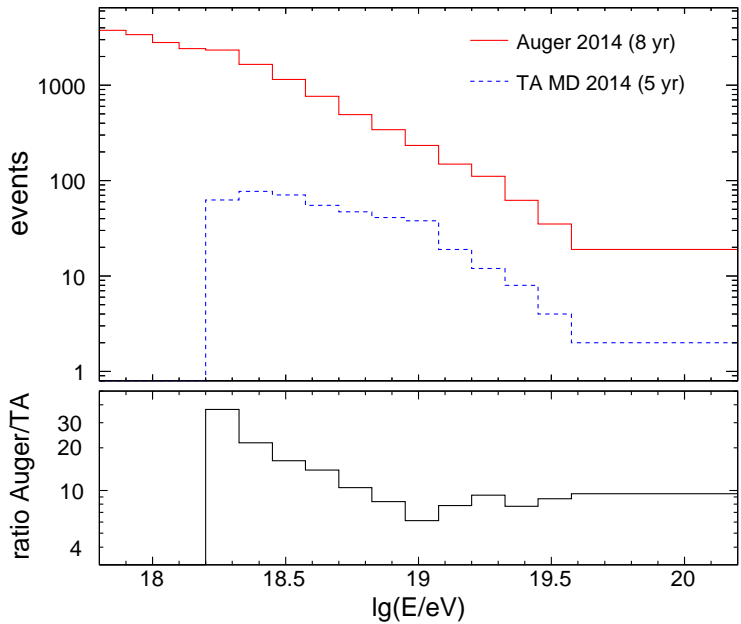}
\caption[events]{Number of selected events for the Auger (solid red
  line) and TA (blue dashed line) analyses. The ratio of events is
  given in the lower panel.}
\label{fig_events}
\end{figure}

\begin{figure}[!t]
\centering
\includegraphics[width=0.48\linewidth]{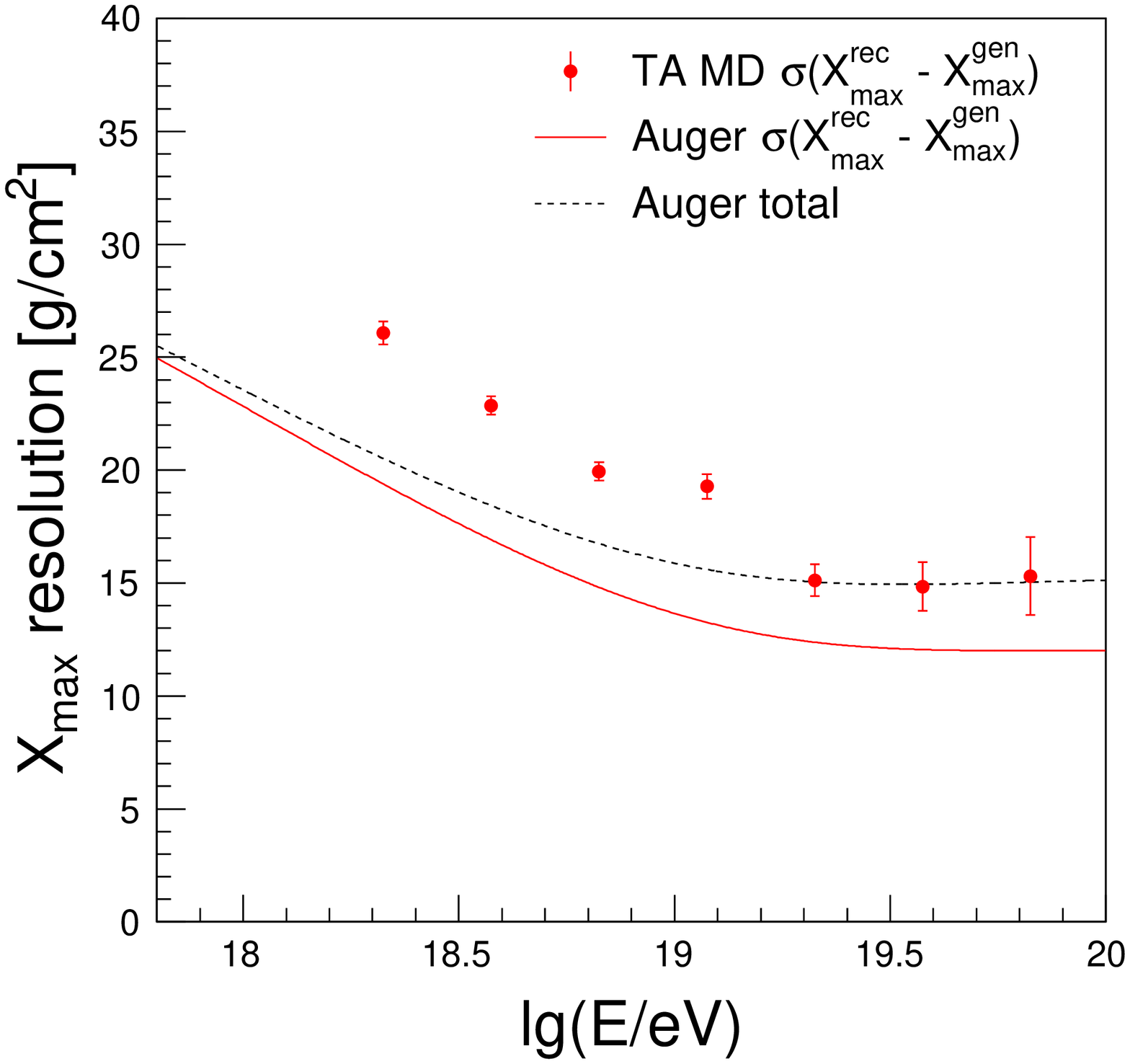}
\includegraphics[width=0.48\linewidth]{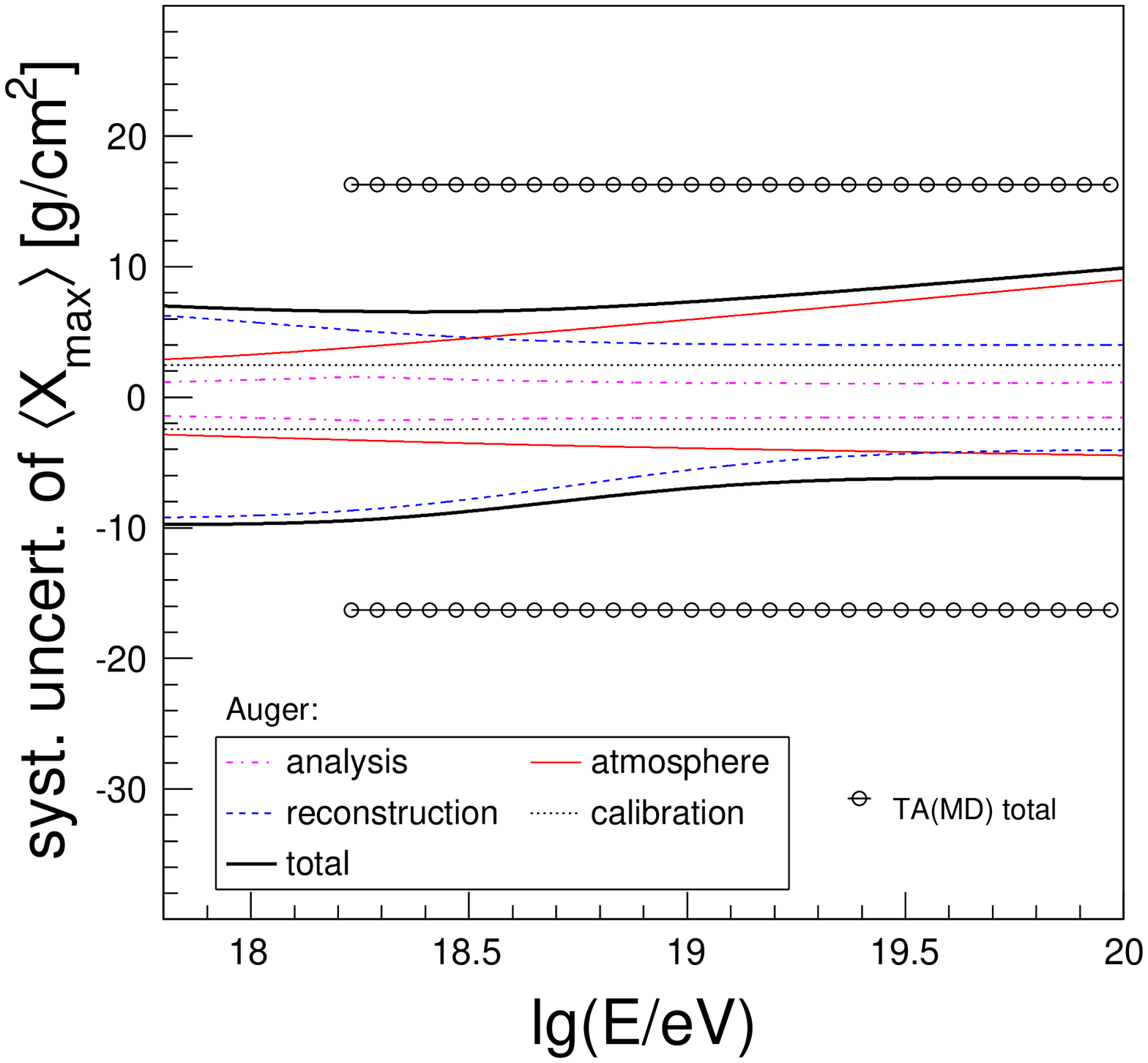}
\caption[events]{\Xmax resolution (left) and systematics of the \Xmax
  scale (right) for the Auger and TA analyses.}
\label{fig_resoAndSys}
\end{figure}

The number of events used for this comparison presented here
is shown in Fig.~\ref{fig_events} and the \Xmax-resolution of the two
experiments is presented in Fig.~\ref{fig_resoAndSys}.
As can be seen, the resolutions after cuts are comparable
but it is worthwhile noting that the
resolution quoted for the MD does not contain effects from the detector
calibration and atmospheric monitoring.
The systematic uncertainties on the \Xmax scale are compared in the
right panel of Fig.~\ref{fig_resoAndSys} and they are $\leq 10$~\gcm
and 16~\gcm for the Auger and TA analyses respectively.

\begin{figure}[!t]
\centering
\includegraphics[width=\linewidth]{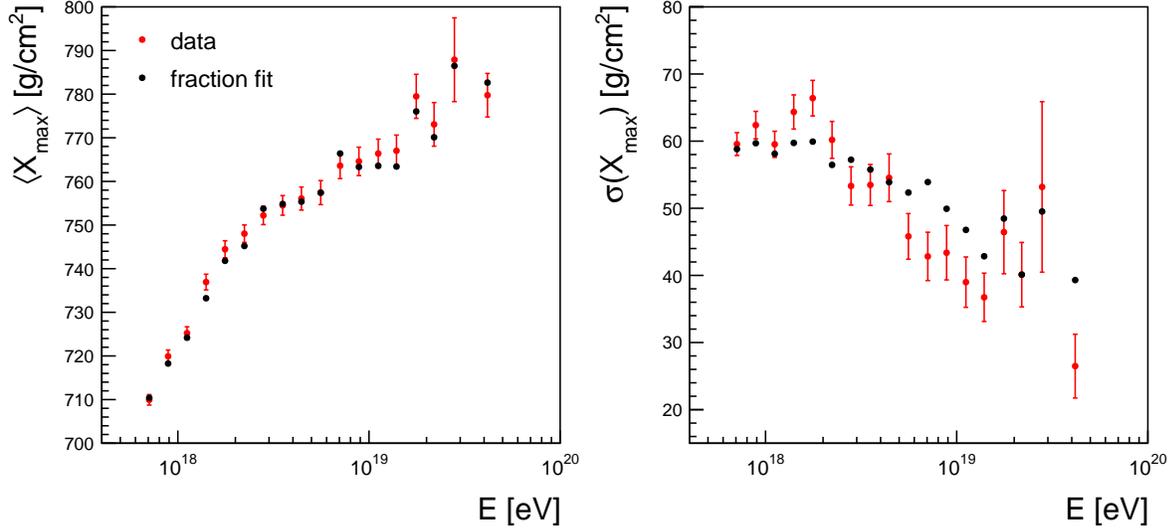}
\caption[qgsjet]{Moments of the fitted \Xmax distributions using
  \QgIIOld (black markers) and \Xmax moments measured by the Pierre
  Auger Collaboration (red circles with statistical error bars), see text.}
\label{fig_fitting}
\end{figure}
\section{Analysis}
Due to the different analysis approaches of the TA and Pierre Auger
Collaborations it is not possible to compare the published values of
\meanXmax and \sigmaXmax directly. Whereas the moments of the \Xmax
distribution published by the Pierre Auger Collaboration are close to
the true  moments (moments of $f_\text{true}$ in
Eq.~\eqref{eq:measurement}), the TA collaboration published the
\meanXmax folded with the effects of the detector response and
reconstruction (moments of $f_\text{obs}$ in
Eq.~\eqref{eq:measurement}).

The relation between
the true and observed \Xmax distribution is
\begin{equation}
f_\text{obs}(X_\text{max}^\text{rec}) =
\int_0^\infty \!\!\!
f_\text{true}(\Xmax)\; \varepsilon(\Xmax)\; R(X_\text{max}^\text{rec} - \Xmax)\;
\dd\Xmax,
\label{eq:measurement}
\end{equation}
i.e., the true distribution $f_\text{true}$ is deformed by the detection
efficiency $\varepsilon$ and smeared by the detector resolution $R$
that relates the true \Xmax to the reconstructed one,
$X_\text{max}^\text{rec}$.

To be able to perform nevertheless a comparison of the two results, we
need to establish what $\langle X_\text{max}\rangle_\text{obs}$ would look like in the TA
detector given the \Xmax distribution measured by Auger.  For this
purpose, we convolute a parametric description of $f_\text{true}$ that
is based on the Auger data with the TA detector simulation and apply
the same reconstruction and analysis chain used for the TA data to
this simulated data set (see~\cite{Hanlon:2013dra} for a previous
description of this method).

Technically, the parametric description of the \Xmax distribution is
realized by providing a set of composition fractions as a function of
energy that describe the \Xmax distributions measured by Auger.  These
fractions are obtained as described in~\cite{Aab:2014aea} by a
log-likelihood fit of templates of \Xmax distributions for different
nuclear primaries as predicted by air shower simulations using a
particular hadronic interaction model. It is worthwhile noting that
the detector acceptance and resolution at a given primary energy
depend mainly on \Xmax itself and only weakly on the primary particle
type or hadronic interaction model via the invisible energy.  Therefore, for
the analysis presented here, it is only important that the resulting
composition mix describes the data well and not which fractions of
primaries are needed or which hadronic interaction model is used to
obtain the model of the 'true' \Xmax distribution.

Here we used \QgIIOld~\cite{Ostapchenko:2010vb} and a mix of four
primary particles (proton, helium, nitrogen and iron) to obtain a
model of true \Xmax distribution based on the Auger data.  \QgIIOld is
not included in the set of models studied by the Pierre Auger
Collaboration to infer the primary composition~\cite{Aab:2014aea}
because it gives a worse description of LHC data than the re-tuned
version \QgIINew~\cite{Ostapchenko:2013pia}. However, with neither
version of \textsc{QGSJetII} it is possible to find a composition mix
that gives a perfect description of the \Xmax distributions measured
by Auger. The first two moments of the best fits with \QgIIOld and the
Auger data are shown in Fig.~\ref{fig_fitting}. As can be seen, there
is a good agreement regarding \meanXmax, but there are deviations
between the fitted and observed width of the distribution.

Ideally, this analysis should be performed with a combination of
composition and hadronic interaction model that fits the Auger data
well, such as \Sibyll~\cite{Ahn:2009wx} or \Epos~\cite{Pierog:2013ria}
(see discussion in~\cite{Aab:2014aea}). However, due to the lack of
large air shower libraries other than \QgIIOld within the TA
Collaboration, we performed the analysis with this model for practical
reasons. Since the deviations between the moments of the data and the
ones of the fitted distributions are on average at the 5~\gcm level,
this approach is expected to give only a small bias in the predicted
observed distributions.

\begin{figure}[!t]
\centering
\includegraphics[width=0.75\linewidth]{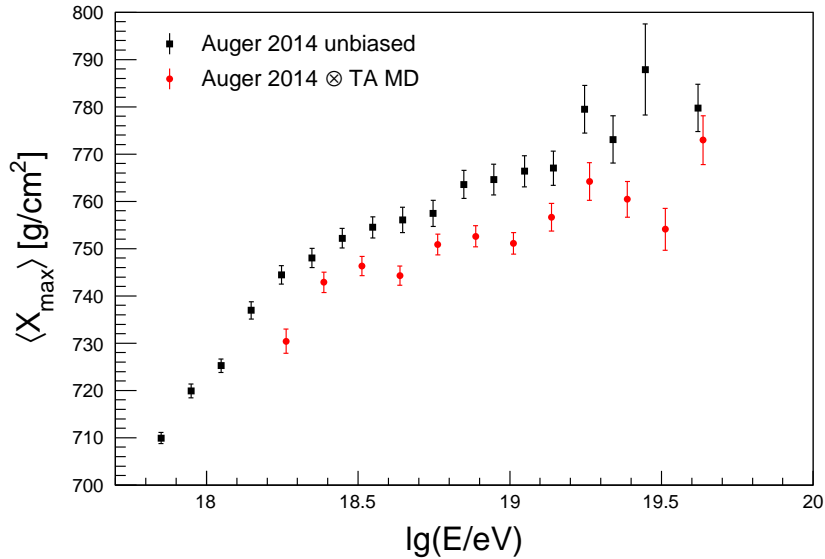}
\caption[dataInMD]{Effect of the MD detector acceptance on \Xmax.
  The \meanXmax of an \Xmax distribution
  describing the Auger data before and after the MD acceptance are
  shown as solid squares and circles respectively. The error bars denote
  the statistical uncertainties of the Auger result in case of the
  squares and the statistical uncertainties due to the limited
  MC statistics in the case of the circles.}
\label{fig_dataInMD}
\end{figure}

In detail, the analysis proceeds as follows: the composition mix is
processed using the Telescope Array hybrid reconstruction analysis
software.  Showers are generated by CORSIKA and the trigger response
of the surface detector is simulated.  The generated longitudinal
shower profile is fitted to a Gaisser-Hillas function to determine the
shower parameters  and a profile based on these parameters is
generated. The TA fluorescence detector response including
atmospheric, electronics, and geometrical acceptance is then simulated.
Subsequently the event geometry is fitted via the fluorescence profile
and the shower-detector plane is measured.  A fit to hybrid shower
geometry is performed which combines the timing and geometric center
of charge of the surface detector array, with the timing and geometry of the
fluorescence detector that observed the event. This step is what makes
the event a “hybrid event”. If either the surface or fluorescence
detector fail to trigger in an event, it is not processed any further,
otherwise the shower profile is fitted via a reverse Monte Carlo
method where the atmosphere, electronics, and geometrical acceptance
of the shower are fully simulated.

The resulting effect of the folding of protons and the parametric
Auger distributions with the TA detector response, reconstruction and
analysis on the \meanXmax of Auger is shown in Fig.~\ref{fig_dataInMD}.
As can be seen, the observed mean is smaller than the unbiased mean.

\begin{figure}[!t]
\centering
\includegraphics[width=0.9\linewidth]{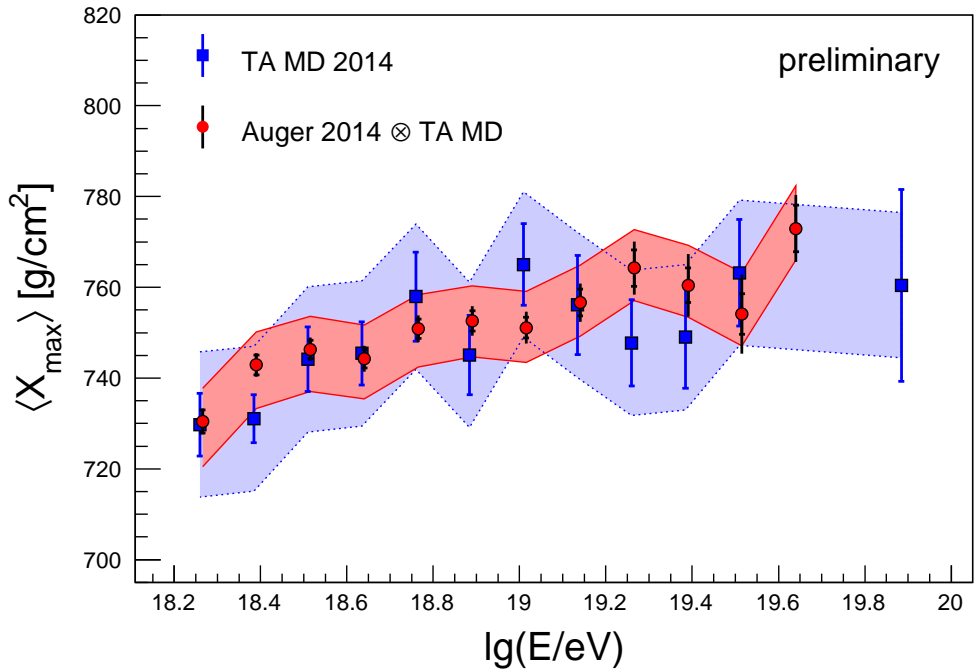}
\caption[dataInMD]{Comparison of \meanXmax as measured with the MD of
  TA (blue squares) and the \meanXmax of the Auger data folded with
  the MD acceptance. The data points were slightly shifted
  horizontally for better visibility. In the case of the Auger points (red circles),
  the inner error bars denote the statistical uncertainty of the
  measurement and the total error bar also includes  contributions from
  the limited statistics of simulated events used for the folding. The colored
  bands show the systematic uncertainties of the \Xmax scales of each experiment.}
\label{fig_result}
\end{figure}

\section{Results and Discussion}
The~\meanXmax as measured by TA using the MD
fluorescence telescope and the Auger result folded with the TA
acceptance are shown in Fig.~\ref{fig_result}.
Their compatibility is quantified with a bin-by-bin comparison
excluding the highest-energy data point of each experiment which
are at different energies. Using only the statistical uncertainties
yields a $\chi^2/$Ndf of 10.7/11 with $P(\chi^2\ge10.7|11) = 0.47$. The average difference
of the data points is
$(2.9 \pm 2.7\;(\text{stat.}) \pm 18\;(\text{syst.}))$~\gcm with
a $\chi^2/$Ndf of 9.5/10 ($P=0.48$).
It can be concluded that the two
data sets are in excellent agreement, even without accounting for the
respective systematic uncertainties on the $\Xmax$ scale.  However, in
the present study we did not take into account a possible difference in the
energy scale of the two experiments. The comparison of the energy
spectra at the ankle region suggests that the energy scale of TA is
about 13\% higher than the one of the Pierre Auger
Observatory~\cite{spectrumUHECR14}. But since the elongation rate of
the folded Auger data is small ($\sim 19$~\gcmdec), the effect of such
an energy shift on the comparison is expected to be at the level of a
few \gcm. For a more precise evaluation it is required to take into
account the energy dependence of the acceptance of TA. Nevertheless,
it is to be expected that the increased difference between the two
data sets once the energy scale shift is taken into account will be
much smaller than the systematic uncertainties on the \Xmax scale of
$\leq 10$~\gcm and 16~\gcm for the Auger and TA analyses respectively.

\section{Conclusions}
In this paper we presented a comparison between the data on \meanXmax
as measured by the Pierre Auger and Telescope Array Collaborations. An
adequate comparison was achieved by taking into account that the
\meanXmax published by Auger is corrected for detector effects,
whereas the \meanXmax published by TA includes detector
effects.  In the future, we intend to improve the parametric
description of the Auger \Xmax distributions and the
evaluation of the effect of the relative energy scale uncertainty. Nevertheless, from the preliminary
comparison presented here we conclude that the data of the two
observatories are in good agreement.

\end{document}